\documentclass[sigconf,screen]{acmart}

\setcopyright{acmlicensed}
\copyrightyear{2025}
\acmYear{2025}
\acmDOI{XXXXXXX.XXXXXXX}
\usepackage[utf8]{inputenc} 
\usepackage[T1]{fontenc} 

\acmConference[]{}{}{}
\acmISBN{978-1-4503-XXXX-X/2025/06}

\def\MM{R1}
\def\RBR{R2}
\def\BC{R3}
\def\EBH{R4}
\def\JH{R5}

\def\B{P1} 
\def\A{P2} 
\def\M{P3} 
\def\R{P4} 

\def\bEi{P6}

\def\uVs{P7}
\def\Utz{P8}

\def\gBt{\A}

\def\Waj{P19}

\usepackage{microtype}
\begin{document}

\author{Manas Mhasakar}
\affiliation{%
  \institution{Ka Moamoa Lab}
  \department{Georgia Institute of Technology}
  \city{Atlanta}
  \state{GA}
  \country{USA}
}
\email{manasmhasakar@gatech.edu}
\author{Rachel Baker-Ramos}
\affiliation{%
  \institution{Ka Moamoa Lab}
  \department{Georgia Institute of Technology}
  \city{Atlanta}
  \state{GA}
  \country{USA}
}
\email{rachelbaker@gatech.edu}
\author{Benjamin Carter}
\affiliation{%
  \institution{Ka Moamoa Lab}
  \department{Georgia Institute of Technology}
  \city{Atlanta}
  \state{GA}
  \country{USA}
}
\email{bcarter81@gatech.edu}
\author{Evyn-Bree Helekahi-Kaiwi}
\affiliation{%
  \institution{Lau By Keha LLC}
  \city{Phoenix}
  \state{AZ}
  \country{USA}
}
\email{laubykeha@gmail.com}
\author{Josiah Hester}
\affiliation{%
  \institution{Ka Moamoa Lab}
  \department{Georgia Institute of Technology}
  \city{Atlanta}
  \state{GA}
  \country{USA}
}
\email{josiah@gatech.edu}


\title[{\textit{Kumu} Perspectives on LLMs for Culturally Revitalizing CS Education in Hawaiian Schools}]{“I Would Never Trust Anything Western”: Kumu (Educator) Perspectives on Use of LLMs for Culturally Revitalizing CS Education in Hawaiian Schools}



\begin{abstract}
As large language models (LLMs) become increasingly integrated into educational technology, their potential to assist in developing curricula has gained interest among educators. 
Despite this growing attention, their applicability in culturally responsive Indigenous educational settings like Hawai`i’s public schools and Kaiapuni (immersion language) programs, remains understudied.
Additionally, \textit{`Ōlelo Hawai`i}, the Hawaiian language, as a low-resource language, poses unique challenges and concerns about cultural sensitivity and the reliability of generated content.
Through surveys and interviews with \textit{kumu} (educators), this study explores the perceived benefits and limitations of using LLMs for culturally revitalizing computer science (CS) education in Hawaiian public schools with Kaiapuni programs.
Our findings highlight AI's time-saving advantages while exposing challenges such as cultural misalignment and reliability concerns. 
We conclude with design recommendations for future AI tools to better align with Hawaiian cultural values and pedagogical practices, towards the broader goal of trustworthy, effective, and culturally grounded AI technologies.

\end{abstract}


\begin{teaserfigure}
  \includegraphics[width=\textwidth]{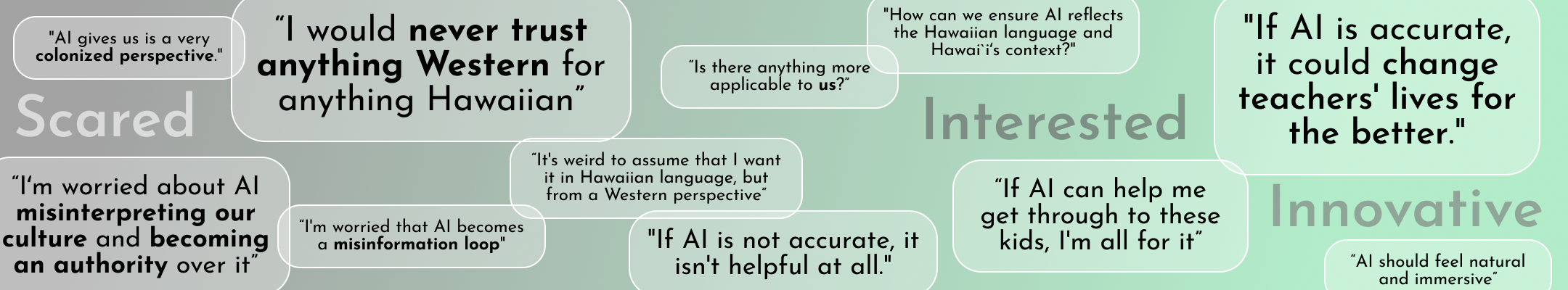}
  \caption{Teacher quotes from surveys and interviews}
  \Description{A gradient background transitioning from gray on the left to green on the right, with the words 'Scared' far left, 'Interested' near the center-right, and 'Innovative' far right, displayed in varying sizes. Several quotes are overlaid across the image, with the large quotes 'I would never trust anything Western for anything Hawaiian' at the top left and 'If AI is accurate, it could change teachers' lives for the better' at the top far-right. The remaining quotes vary in size and are distributed across the top and bottom of the graphic.}
  \label{fig:teaser}
\end{teaserfigure}
\maketitle

\section{Introduction}

The integration of computer science (CS) education into K–12 curricula has become a global priority, aiming to equip students with essential digital-age skills. For many school systems, CS education has become a requirement due to state-level mandates to meet CS standards \cite{Act51, Act158}. 
These mandates, while well-intended, exacerbate disparity issues with under-resourced schools, which can lack the resources for CS curricula development and the educators with the technical expertise to teach CS 
\cite{Nguyen2020CSHawaii}.
In Hawai`i, this issue is compounded by its unique context: Public schools in Hawai`i require \textit{“Hawaiian language, culture, and history"} to be an \textit{"integral part of Hawai`i’s education standards”} in all subjects \cite{OHEDeliveryPlan}. 
Additionally, 22 of these public schools have Hawaiian language immersion programs, known as Kaiapuni \cite{HIDOE_Kaiapuni}, which deliver instruction that is both culturally relevant and primarily conducted in \textit{`Ōlelo Hawai`i} (The Hawaiian Language) \cite{Nguyen2020CSHawaii}. 
These schools play a pivotal role in the revitalization of Hawaiian culture and \textit{`Ōlelo Hawai`i}, which have been historically marginalized through colonization. 
Unfortunately, the impacts of colonization affect educators, as they face the challenge of creating culturally relevant lessons for a culture that was actively exploited and suppressed and, for Kaiapuni teachers, in a language that was outlawed. Finally, there will soon be the additional challenge within this context: Starting 2025, all schools in Hawai`i will be required to teach Computer Science \cite{Act158}.

Alongside the growth in CS education, curriculum development tools are evolving, with AI technologies, particularly large language models (LLMs), becoming increasingly integrated into educational settings.
In Hawai`i, while these tools could potentially aid in curriculum development and address the shortage of teaching resources, cultural alignment and sensitivity must be part of an AI solution. 
This presents unique challenges, as \textit{`Ōlelo Hawai`i} is a low-resource language, with limited digital support for tasks \cite{ijcai2023p0685}.
Furthermore, many Indigenous researchers are skeptical about AI’s capacity to respect Indigenous nuances \cite{CardonaRivera2024, DBLP:journals/corr/abs-2009-14258, Sangma2024}, intensifying fears of digital colonization, where dominant cultures overshadow Indigenous ones\cite{RePEc:sae:envira:v:51:y:2019:i:7:p:1424-1441, doi:10.1177/0306396818823172}. The risk of AI tools misrepresenting or diluting Indigenous knowledge systems necessitates a careful and culturally sensitive approach to their adoption \cite{Lewis2024, DBLP:journals/corr/abs-2009-14258}.

The following paper presents initial research, conducted through surveys and interviews with educators from schools with Kaiapuni programs, on exploring existing curricula development processes, the integration of AI into educational practices, and educators perceptions of its opportunities and limitations."
We present findings paired with design recommendations to guide future AI tools to align with Hawaiian cultural values, pedagogical practices, and language nuances. 
This work contributes to the broader conversation on designing culturally responsive AI systems in education, especially for low-resource languages, and offers actionable insights for developing more effective, culturally grounded AI technologies that educators can trust.

\section{Related Works}
Despite growing interest in AI and LLMs, research exploring how these tools align with the linguistic and cultural priorities of Indigenous communities remains limited. Specifically, no prior studies to our knowledge, have evaluated the reception of these systems in Hawai`i, including Hawaiian public schools with Kaiapuni programs, where the preservation of \textit {`Ōlelo Hawai`i} and cultural identity are paramount. Understanding how LLMs can support culturally responsive pedagogy in low-resource language environments is crucial for advancing AI solutions that honor local values, strengthen Indigenous knowledge systems, and fully meet educators’ pedagogical objectives.

\subsection{Educators Views on AI in Education}
As LLMs continue to advance, educators are exploring their potential to automate routine tasks, create culturally relevant content, and assist in lesson planning. Studies demonstrate that these models can bridge gaps between curriculum goals and effective classroom delivery by generating curriculum-aligned lesson plans \cite{Fan_2024,2023LeveragingGenerative,ContentAnalysisofLessonPlans,karaman2024lessonplans,mahtoum2024investigating}, while also supporting adaptive learning and enhancing student engagement \cite{10.1007/978-981-97-1814-6_18, 10.1145/3674829.3675082}. Early adopters feel that AI can serve as a valuable brainstorming partner for innovative teaching strategies while reducing educators' workloads \cite{KaplanRakowski2023, 2023LeveragingGenerative, UseofChatGPTforLessonPlanning, HADIMOGAVI2024100027, 10490240}. Despite these advancements, research has shown that educators face significant barriers that impede the seamless adoption of AI. A primary issue is the need to ensure that AI-generated content is accurate, unbiased and suitable for diverse learners, as LLMs are prone to hallucinations and may produce inaccurate, biased, or culturally insensitive material. \cite{dornburg:hal-04558217, naous2024havingbeerprayermeasuring, 10.1093/pnasnexus/pgae346, Liu+2024, sahoo-etal-2024-addressing}. This creates an extra burden for educators, requiring them to review and refine AI-generated outputs to ensure accuracy. Teachers have also raised concerns about data privacy, algorithmic biases, and ethical implications, particularly the risk of over-reliance on AI undermining students’ critical thinking skills \cite{Staake2023,KOPCHA20121109, Akgun2022}. Barriers such as insufficient training, limited resources, and time constraints further hinder the effective integration of AI in education, leaving many educators uncertain about navigating these emerging tools and adapting them to their teaching practices \cite{KASNECI2023102274, Pappa2024, 10.1016/j.chb.2012.10.017, Hsu2016, Moura2024, 10.3389/fpsyg.2024.1468900, articleUygunDerya}. Addressing these challenges requires comprehensive teacher training, clear ethical frameworks, and institutional support \cite{Chounta2022, 10.1145/3491101.3519866} to ensure the responsible and effective use of AI tools in education.

\subsection{Culturally Responsive Pedagogy in  Hawaiian Education}
Culturally responsive pedagogy in Hawai`i is anchored in place-based and `āina-based education\footnote{`Āina means "that which sustains us," referring to place as more than land, but also the people, culture, creatures, and environment.}, which ties learning activities to local environments, community knowledge, and Indigenous epistemologies \cite{NgOsorio2023, harada2016placebased, Porter2025}.
Previous research has consistently shown that weaving students' cultural identities, histories, and language into curricula enhances engagement and academic outcomes, and students develop a deeper attachment and sense of responsibility toward their communities \cite{kanaiaupuni2010culture, taira2023ma, roberts2017placebased, kuwahara2013impacts, Ledward2008}.
Furthermore, culturally relevant computing research \cite{Hoffman2023,10.1145/3653666.3656066, 10675788} and culturally responsive teaching studies \cite{10.1145/3500868.3559473} underscore the importance of involving Indigenous communities in designing computer science and other subject curricula. 
Such collaborations ensure that curricula are not only aligned with local epistemologies and modern educational needs but also effectively address the specific and diverse needs of Indigenous students, thereby fostering equitable and impactful learning environments \cite{Canevez21102022,10.1145/1753326.1753522}.
By centering education on Indigenous knowledge systems that incorporate community expertise and Hawaiian values, like \textit{mālama `āina} (to care for the land), Hawai`i’s educators can create a sustainable and inclusive model for teaching and learning that resonates with both cultural heritage and modern educational needs.

\subsection{LLMs for Low-Resource Languages and Indigenous Communities}
The development and deployment of AI systems for low-resource languages and Indigenous communities presents complex technical, cultural, and ethical challenges. 
Large Language Models (LLMs) are primarily trained on dominant languages \cite{hasan2024largelanguagemodelsspeak} and often fail to adequately capture the nuances of smaller language communities, leading to inaccuracies, mistranslations, and cultural misrepresentations \cite{ijcai2023p0685, zhong2024opportunitieschallengeslargelanguage, ahuja2023megamultilingualevaluationgenerative}. 
This issue can be particularly pronounced in educational settings, where such AI tools risk inadvertently perpetuating linguistic and cultural biases \cite{li2024potentialsocietalbiaseschatgpt, naous2024havingbeerprayermeasuring,alkhamissi2024investigatingculturalalignmentlarge}. 
Beyond technical limitations, ethical concerns such as data sovereignty, digital colonization, and the reinforcement of power imbalances are equally critical to address when deploying AI in Indigenous contexts \cite{10.1145/3653666.3656107, Adams03042021}. 
Addressing these challenges requires building technical solutions through an ethical and decolonial lens, leveraging community-led approaches to data gathering and model development, and prioritizing culturally sensitive Natural Language Processing (NLP) techniques \cite{pinhanez2024harnessingpowerartificialintelligence}. 
Crucially, AI solutions must be co-created with Indigenous communities, respecting their values, decision-making structures, and aspirations for self-determination. Such an approach transcends technical optimization, ensuring that emerging technologies foster inclusivity and support the preservation and revitalization of low-resource languages.

\section{Methodology}
For this research, we conducted surveys with 15 educators teaching across K-12 grade levels and semi-structured interviews with 4 educators, all from public schools with Kaiapuni programs. We grounded the survey and interview questions in seven core themes: (1) Cultural Sensitivity and Representation, (2) Data Sovereignty and Ethical Concerns, (3) Workforce and Impact on Educators, (4) Usability and Accessibility, (5) Perceived Opportunities and Risks, (6) Practical Suggestions and Design Recommendations, and (7) Vision and Future Outlook. Additionally, we included questions on demographics, background, and experience with AI, to evaluate how these factors impact each theme. 
The fourth author reviewed the research instruments for cultural sensitivity. 
We then analyzed these using a combination of thematic qualitative analysis and quantitative methods. The first and second authors coded the qualitative data from the interviews and open-ended survey questions, aligning quotes with one of the seven themes. The first and third authors analyzed quantitative data from survey results. Finally, we used a mixed-methods approach to inform the results below. 

\section{Results}
Survey respondents included homeroom teachers, subject-specific instructors, and stand-ins like parents or community members, with most reporting over 10 years of teaching experience, reflecting their extensive familiarity with the challenges and opportunities in Hawaiian education.
While many identified as \textit{Kanaka Maoli} (Native Hawaiian), others, including those identifying as Asian, White, or Pacific Islander, predominantly described themselves as \textit{Kama`āina} (Local), reflecting their long-term residence and deep social ties to Hawai`i, despite differing ethnic backgrounds. 
This underscores both the cultural diversity and the shared sense of belonging within the Hawaiian educational community.

\subsection{Cultural Sensitivity and Representation}
Survey and interview responses highlighted the importance of culturally representative education in Hawai`i, with nearly 75\% of participants responding "\textit{very important}" or "\textit{extremely important}" to the survey question, "How important is it to integrate Hawaiian cultural values and knowledge into your CS teaching?."
Educators explained how cultural representation in lessons necessitates the integration of local references and \textit{mo`olelo} (stories and legends).
\R\ referred to this approach as "\textit{`āina}-based learning," which incorporates "\textit{Hawai`i values and stories and ways of doing things}," as well as "\textit{Hawaiian-based science}." \R\ went on to explain the interconnected nature of this approach, as "\textit{`āina is connected to the culture. You can't separate them.}"
\M\ explained how this is a continuous thread in the students' educational experience, from morning \textit{oli} (songs), to local \textit{mo`olelo} in the classroom lessons, to frequent field trips to community sites where "\textit{members of the community will share their stories}," which is especially engaging for his students, because "\textit{sometimes it's their uncle or their auntie}."
As \B\ described, "\textit{students are really engaged when they're learning about things that they know of and that they see on a daily basis}."

\B\ has used AI tools to create culturally relevant materials, but found the output lacking in cultural and regional accuracy, noting that "\textit{even when you ask it to be specific, it kind of just moves the words around.}" 
From her collaboration with \textit{`Ōlelo Hawai`i} speakers, she saw how, with the exception of Gemini \cite{Gemini}, the AI "\textit{translation has not been very accurate either}," producing outputs that have to be revised by hand.
\B\ also shared a compelling personal example of cultural misrepresentation, describing how ChatGPT portrays colonization in a \textit{"positive light"}, glorying \textit{"people who came in and colonized"} Hawai`i. 
She cautioned that such portrayals are detrimental to cultural revitalization efforts and could harm students by presenting a skewed narrative of Hawai`i’s history, warning that this could cause \textit{“a whole generation"} to view their own culture through a colonized lens.
Similarly, \R\ felt that she "\textit{really had to force [AI] to use Hawaiian [terms]}" and was disappointed with AI outputs defaulting to stereotypical wording, like "\textit{in the lush tropical paradise of Hawai`i}."
\A, a technology educator who also collaborates with \textit{`Ōlelo Hawai`i} speaking educators, found some success in culturally and linguistically representative outputs, but only after modifying the prompt to ask for output "\textit{from an Indigenous Hawaiian epistemology}." 
She reflected on the disconnect within AI tools that necessitated this extra step, wishing that the tool would automatically connect language and culture: "\textit{It shouldn't have to be requested separately. It's kind of weird to assume that I want it in Hawaiian language, but I want it from a Western perspective}."
Notably, many educators, regardless of their level of experience with AI, explicitly cited concerns about cultural misrepresentation in AI outputs and its potential consequences \textit{(see Section \ref{OpAndRisks} Perceived Opportunities and Risks)}.

\subsection{Data Sovereignty and Ethical Concerns}
The importance of source validity and transparency emerged across surveys and interviews as another critical aspect of culturally responsive education. Educators shared how the source for legitimate and accurate local \textit{mo`olelo} and Hawaiian knowledge is primarily verbal story telling within the community, along with a few Hawaiian-made and -owned databases and archives.
\M\ teaches at a rural school in Maui and, while he is \textit{kama`āina} (local) and knows broader-reaching Hawaiian \textit{mo`olelo}, he primarily uses \textit{mo`olelo} specific to the town, as students respond best to this. His source of local \textit{mo`olelo} is his wife, who grew up in the school's town and "\textit{knows all the stories that pertain to this environment and to this community}." \B, who is \textit{malihini} (non-Hawaiian and non-local), shared that she and fellow teachers often got "\textit{the information from a Kupuna }(Elder)," who "\textit{already had that knowledge}" and passed it down to the community.
\R, a \textit{kama`āina} librarian, uses a small number of databases that have been validated as Hawaiian-owned and -made, but are time-consuming to navigate \textit{(see Section \ref{Workforce} Workforce and Impact on Educators)}.

Educators cited source validity and transparency, or lack thereof, as a primary reason they have hesitated to use AI tools.
In response to the survey question "In what other ways would you like AI tools or existing tools to be adapted to better meet your needs as an educator?", \gBt\ shared that they feel "\textit{hesitant to use AI tools because I am unsure about where the information is coming from}." 
\Waj\ echoed this in their response to the same question: "\textit{I want to know where the AI is grabbing the info from}", highlighting why validated and accurate sources are so important for educators. 
\gBt\ reiterated concerns about AI transparency conveying when it does lack accurate information, as "\textit{a response is usually generated even if the AI is not confident in the content, and that is concerning}." 
A few educators also expressed concern about the inherently Western bias embedded in AI-generated outputs due to their training on datasets dominated by colonized perspectives. 
\R\ explained that she is concerned with "\textit{what information is 'training' the AI, how that information is cited/protected, and what does the AI when it does not know something}." \B\ highlighted how these datasets often fail to reflect local knowledge or cultural context : \textit{"A lot of the books and texts that have been scanned into the AI come from the mainland, which is a very colonized thought. The information that we're getting from the AI is shaped by this colonized perspective."}
\B\ further emphasized the disconnect between these outputs and the specific needs of Hawaiian schools:\textit{
"I think that that's been difficult because I don't know if AI understands the difference between what responses would work best for where we live."}
Many educators shared these concerns surrounding misinformation \textit{(see Section \ref{OpAndRisks} Perceived Opportunities and Risks)}.

\subsection{Workforce and Impact on Educators}\label{Workforce}
In response to questions on concerns about AI impacting the role of teachers, educators brought up the time-consuming nature of generating curricula for culturally relevant and immersion education, and the need for more tools. 
As an example of this high effort, \B, who teaches in English, but integrates \textit{`Ōlelo Hawai`i} into her lessons, explained that creating lessons "\textit{that were Hawaiian-based with \textit{`Ōlelo Hawai`i} included}" took around "\textit{6 weeks to build out with other teachers involved}."
Integrating cultural representation into Computer Science exacerbates this time-consuming process. The survey revealed that a majority of participants (9 participants) experienced the search for culturally relevant CS materials in \textit{`Ōlelo Hawai`i} as "\textit{extremely challenging}," with a further 5 finding it "\textit{very challenging}." 

Within this context, educators \B\ and \R\ shared the sentiment that AI would not have the capabilities to diminish the role of educators, as the greater need is to reduce the burden on teachers, especially \textit{Kanaka Maoli} (Native Hawaiian) teachers. \M\ shared that he was slightly worried but if AI is "\textit{another tool to teach and it doesn't replace the actual teacher, then I'm all for it}." \textit{(see Section \ref{OpAndRisks} Perceived Opportunities and Risks)}. \A\ in their role as a technology educator highlighted that the challenge lies in educators' familiarity with technological tools rather than their expertise. "\textit{There’s so much anxiety tied to technology}," they observed, adding, "\textit{I talk to teachers and state professionals who don’t feel confident in their abilities, even though they’re experienced educators.}" They concluded, "\textit{One of the big barriers is that people need to feel confident in using technology, it’s not about their professional capacity but about familiarity with the tools.}"

\subsection{Usability and Accessibility}
The survey highlighted growing AI awareness among educators: 4 were aware of AI but had not used it, 6 had limited knowledge and occasional usage, and 5 were occasional users of AI tools, with several respondents citing a lack of training and understanding of AI as main barriers.
While the Hawai`i Department of Education (HIDOE) has taken a major step by approving access to AI platforms \cite{HIDOE_AI} like Google AI’s Gemini and NotebookLM, a significant confidence gap emerged when we asked educators, "How comfortable are you with creating prompts for AI tools to create teaching materials?" A considerable portion felt either "\textit{not comfortable}" (4 respondents) or "\textit{slightly comfortable}" (5 respondents). Only 2 respondents reported feeling "\textit{moderately comfortable}" or "\textit{very comfortable}." These findings align with broader trends indicating insufficient AI training among educators \cite{weiner2025survey, edweek2024survey, meyer2024aiuse, schwartz2024disadvantage}.
Qualitative data from our surveys and interviews reinforced this need for training, with \bEi\ stating “\textit{I just need more training and support on how to use AI}” and \M\ sharing that he "\textit{would love to learn how to use it and use it effectively.}"
\A\ also expressed that "\textit{when someone doesn’t understand how to phrase the prompts, they might create outputs that are not as appropriate or culturally aligned}."
Current training opportunities, limited to online courses, have proven to be insufficient for many educators, highlighted by \M, who expressed frustration with self-paced online courses, \textit{"I prefer one-on-one or interactive training..someone to keep me on my toes..rather than just clicking on links to watch 6 hour long videos or read materials."}
%
%
%
\subsection{Perceived Opportunities and Risks}\label{OpAndRisks} 
In response to the open-ended survey question "What's 1-3 words to describe how you feel about Artificial Intelligence (AI)?" the 2 most commonly written words were "scared" and "interested," closing followed by "innovative." 7 respondents expressed an overall positive outlook, using descriptors like “innovative” and “efficient.” Conversely, 3 conveyed negative sentiments such as “scared” and “intimidating,” while 5 reflected mixed feelings, highlighting terms like “uneasy,” “unsure,” and “anxious.” These responses showed a range of emotional reactions to AI’s potential role in education.

Educators spoke of the risks of misinformation and cultural misrepresentation going unnoticed, impacting the curriculum and students' education.
\R\ illustrated how users must diligently "\textit{check the AI's work and have a functional understanding of information science to do so - which is not a universal condition}." \R\ also explained the risk of "\textit{misinformation making it into the curriculum} when educators use AI "\textit{without mitigating its flaws}."
\Waj\ shared this concern around misinformation \textit{becoming an inaccurate source}, expressing worry that AI would not "\textit{find the right info}," and would instead "\textit{[make] it up to fill gaps}."
This risk goes beyond misinformation, also surfacing issues of digital colonization, with \Waj\ "\textit{worried about AI misinterpreting our culture and then becoming some kind of authority over it}."
\A\ also shared that "\textit{One of the challenges with AI tools like ChatGPT is that they sometimes generate vocabulary or concepts that don’t align with known Hawaiian sources and there needs to be transparency in where the AI is pulling this information from}."
Educators recognized the potential of AI tools to greatly reduce the time it takes educators to create culturally relevant lessons, as a potential support mechanism for substitutes or non-specialist teachers and, for Kaiapuni teachers, create those lessons in \textit{`Ōlelo Hawai`i}.
\B\ viewed "\textit{getting time back}" as the main benefit of AI, specifically "\textit{being able to see how much time you can cut out of your preparatory work has been huge}."
\R\ spoke about the potential of AI tools, if validated as a reliable source, to help educators \textit{"gain cultural and language contextual understanding, without burdening working \textit{kumu} (teachers),"} allowing all teachers to "\textit{more effectively teach place-based, relevant, [and] appropriate content}."
\subsection{Practical Suggestions and Design Recommendations}
The survey asked educators to rank the most helpful features in an AI tool designed for Hawaiian Immersion education. Top priorities included the easy creation of culturally relevant lesson plans and high-quality \textit {`Ōlelo Hawai`i} language support, both of which consistently ranked among the top 3, followed closely by integration of place-based knowledge and intuitive, easy-to-use interfaces.
Educators provided a range of suggestions for AI tools, from specific features to database organization, in the interview and survey.
Specific features focused on artifact generation, such as "\textit{illustrations for \textit {`Ōlelo Hawai`i} children's books}" (\Utz) and "\textit{presentations to teach curriculum}" generated from a research paper "\textit{to expedite the process of teaching from the paper}" (\uVs). 
\R, a librarian, offered suggestions on database and search mechanics that could align with Hawaiian values and perspectives. She gave the example of Western authorship, which typically assumes the person who wrote a concept down is who should have credit for it. In contrast, Hawaiian \textit{mo`olelo} may have been written by one person, but will include where the origination originates from and who has passed the story down. \A\ suggested a long-term goal for AI: "\textit{these systems should inherently recognize that if you’re asking for a Hawaiian language lesson, it should default to incorporating Indigenous epistemology rather than needing explicit instructions}." They added that "\textit{the experience should feel natural and immersive, rather than requiring extra effort from educators}."
\subsection{Vision and Future Outlook}
From the survey, 12 of the 15 participants expressed willingness to use AI tools specifically designed for Hawaiian immersion contexts, while 3 were uncertain. This enthusiasm aligns with perceptions of AI’s potential to support culturally relevant education: 8 participants viewed AI’s role as “\textit{very positive},” while 4 described it as “\textit{somewhat positive}.” However, 3 participants expressed a “\textit{somewhat negative}” outlook, reflecting a mix of optimism and caution about AI’s ability to effectively integrate Hawaiian language and place-based knowledge into CS education. Within the short span of this study, it was clear how rapidly views are changing: One survey respondent, \M, who initially wrote  "\textit{scared}" for his perception of AI, shared his shifting perspective in his interview a few weeks later, reporting that he had "\textit{heard so many really good things about AI that I'm down to try it out and for my practice here in the classroom}."


\section{Discussion}
The results emphasize the importance of developing educator tools for Hawaii`s public schools and Kaiapuni programs focused on supporting immersion and culturally relevant computer science education. Alongside this is the critical need for these tools, if AI-based, to be culturally aligned, linguistically appropriate, and user-friendly.
This section outlines key findings and design recommendations, focusing on creating AI systems that prioritize cultural sensitivity and trust, while empowering educators through robust training, localized data, and safeguards against misrepresentation.
\subsection{Designing for Cultural Sensitivity, Transparency, and Trust}
Teachers who had more experience with prompt engineering expressed frustration at AI tools not linking requests for Hawaiian language outputs with culturally representative outputs.
\textbf{An AI tool to support culturally revitalizing education should have default assumptions that align with Hawaiian ways of knowing, rather than Western standards.} Such defaults will help in ensuring that AI-generated content naturally reflects local epistemologies without requiring extensive prompt engineering from the educators.
Our results displayed that very few trusted sources for Hawaiian knowledge are digitized, making educators wary of AI that may draw from questionable or Western-biased training sets given this disparity of training data. For teachers who are \textit{malihini} (non-Hawaiian and non-local), finding and validating Hawaiian sources can be a greater challenge, which \B, a \textit{malihini} teacher, described as "one of the biggest challenges" for her. \textbf{In order for an AI tool to be trusted, it must clearly convey data sources, including metrics teachers use to validate said sources, such as Hawaiian-made and Hawaiian-owned}. By clarifying and citing the origins and credibility of its underlying data, an AI system can build trust and demonstrate its respect for Hawaiian knowledge systems.
Finally, teachers who are \textit{Kanaka Maoli} or \textit{kama`āina} may have the contextual knowledge and resources to validate Hawaiian sources, but relying on them alone can be burdensome. Educators who lack this background and context may be unaware of misrepresentation happening, risking inadvertent dissemination of culturally inappropriate content. \textbf{An AI tool must safeguard against this situation by highlighting knowledge gaps where they exist, thus reducing reliance on individual expertise and preventing the propagation of misleading or culturally insensitive information}. By integrating robust training, localized data, and protective measures against misrepresentation, AI systems can become valuable tools for enhancing culturally responsive education.

\subsection{Empowering Educators Through Training, Localized Support, and Safeguards}
Beyond cultural alignment, a recurring theme in our interactions with educators was that a lack of familiarity with prompt engineering and AI tools hindered effective adoption. \textbf{In order for an AI tool to support educators, it should have robust training resources that include prompt engineering guidance, enabling teachers to generate culturally relevant educational materials.} The tool must also have user-friendly interfaces built for teachers who may not be tech-savvy and helpful features such as intuitive workflows for prompt engineering, clear labeling of references, and visual aids for evaluating content reliability.
Our findings and related works also revealed the importance of place-based learning in creating meaningful, culturally responsive curricula. Teachers use \textit{mo`olelo} that is specific to a school's town and immediate surrounding to create successful, engaging culturally responsive education. Consequently, \textbf{an AI tool to support culturally relevant education in Hawai`i should categorize data by place, link data along relevant \textit{mo`olelo}, and prioritize references from a school's town and surroundings.} Such an approach ensures that the generated content reflects local histories, traditions, and landscapes. 

\section{Conclusion and Future Work}
The findings and design recommendations of this work point to the benefits and challenges of AI for culturally relevant education. 
While these findings can inform how general purpose LLMs handle culturally relevant education, we believe these challenges call for curated AI solutions, developed in collaboration with the communities they serve.
Future work will build upon the research presented here to co-design AI systems suitable for the unique context of Hawaiian public schools, particularly Kaiapuni education. These efforts will prioritize alignment with Hawaiian epistemologies, localized learning, and safeguards against cultural misrepresentation, paving the way for culturally sensitive educational tools.


\bibliographystyle{ACM-Reference-Format}

\appendix

\section{Positionality Statement}
The authors bring diverse perspectives to this exploration of teacher perceptions of AI tools in Hawaiian educational settings, with cultural outsiders—\MM, \RBR, and \BC—approaching the research through their expertise in AI for education and community-engaged design, while insiders—\EBH\ and \JH—offer deep, lived connections to Hawaiian language, culture, and computing practices. This blend of perspectives ensures that the research is grounded in community-driven insights.
\MM\ is an MS-CS student who specializes in AI for education with specific experience in culturally relevant translations. 
\RBR\ is a research scientist with expertise in culturally relevant educational technology, decolonization efforts, and community-engaged design. 
\BC\ is a BS-CS student who brings experience in computer science education. 
\EBH\ is a \textit{Kanaka Maoli} 3rd-generation \textit {`Ōlelo Hawai`i} speaker, providing experience in Hawaiian language and culture. 
\JH\ is a \textit{Kanaka Maoli} professor of computing. He provides guidance on ethical computing research in Indigenous contexts, drawing on expertise in revitalizing computing practices and education.

\section{Survey Questions}
Background:
\begin{enumerate}
  \item What is your current role?
  \begin{itemize}
      \item Homeroom teacher
      \item Subject specific teacher (please specify subjects)
      \item Official Substitute teacher
      \item Stand-in Teacher (Parent, community member, etc who is asked to stand-in)
      \item Administrator
      \item Other (please specify)
  \end{itemize}
  \item How many years have you been teaching?
  \begin{itemize}
      \item Less than 1 year
      \item 1-5 years
      \item 6-10 years
      \item 11-15 years
      \item 16-20 years
      \item Over 20 years
      \item Not an educator
  \end{itemize}
  \item Which grades have you taught/supported? (Select all that apply)
  \begin{itemize}
      \item K-2
      \item 3-5
      \item 6-8
      \item 9-10
      \item 11-12
      \item Other (please specify)
      \item None of the above (not an educator)
  \end{itemize}
  \item Which language medium do you primarily teach in?
  \begin{itemize}
      \item \textit {`Ōlelo Hawai`i} (Hawaiian Language)
      \item English
      \item Both \textit {`Ōlelo Hawai`i} and English
      \item None of the above (not an educator)
  \end{itemize}
  \item What is your level of experience with teaching Computer Science (CS) or technology-related subjects?
  \begin{itemize}
      \item Extensive experience
      \item Some experience
      \item Minimal experience
      \item No experience
  \end{itemize}
  \item What are the main challenges you face when teaching CS? (select all that apply)
  \begin{itemize}
      \item Lack of CS teaching materials in \textit {`Ōlelo Hawai`i}
      \item Limited personal expertise in CS
      \item Difficulty integrating CS with Hawaiian cultural context
      \item Students' varying levels of interest or background in CS
      \item Insufficient time to prepare CS lessons
      \item Limited access to technology resources
      \item Other (please specify):
  \end{itemize}
  \item How important is it to integrate Hawaiian cultural values and knowledge into your CS teaching?
  \begin{itemize}
      \item Extremely important
      \item Very important
      \item Moderately important
      \item Slightly important
      \item Not important at all
  \end{itemize}
  \item How challenging is it to find CS teaching materials in \textit {`Ōlelo Hawai`i} or resources that are culturally relevant?
  \begin{itemize}
      \item Extremely challenging
      \item Very challenging
      \item Moderately challenging
      \item Slightly challenging
      \item Not challenging at all
  \end{itemize}
\end{enumerate}
Familiarity with AI:
\begin{enumerate}
  \item How familiar are you with Artificial Intelligence?
  \begin{itemize}
      \item I am very familiar and use AI tools regularly
      \item I am familiar and use AI tools occasionally.
      \item I have limited knowledge and have used it a few times.
      \item I have heard of AI but have not used it.
      \item I have never heard of AI.
  \end{itemize}
  \item What's 1-3 words to describe how you feel about Artificial Intelligence? 
  \item Have you ever used AI tools (e.g., ChatGPT) to assist with teaching or lesson planning?
  \begin{itemize}
      \item Yes
      \item No
      \item Not sure
  \end{itemize}
  \item Which AI tools have you used? (select all that apply)
  \begin{itemize}
      \item ChatGPT
      \item Gemini
      \item Microsoft Bing AI
      \item Khanmigo
      \item MagicSchool AI
      \item Notebook LM
      \item Other (please specify)
  \end{itemize}
  \item For what purposes have you used AI tools in your teaching?
  \begin{itemize}
      \item Generating lesson plans
      \item Creating culturally relevant teaching materials
      \item Translating content into \textit {`Ōlelo Hawai`i}
      \item Developing CS curriculum content
      \item Assisting stand-in teachers with materials
      \item Creating impromptu activities 
      \item Grading or assessing student work
      \item Other (please specify)
  \end{itemize}
  \item What concerns or barriers prevent you from using them?
  \begin{itemize}
      \item I don’t understand enough about AI tools
      \item I have concerns about cultural misrepresentation
      \item I’m concerned about AI replacing teachers
      \item I’m concerned about AI replacing teachers
      \item I don’t have access to this technology
      \item Other (please specify)
  \end{itemize}
\end{enumerate}
Perspectives on AI in Education:
\begin{enumerate}
  \item How do you feel about the potential of AI to support education in Hawaiian Immersion Language Schools? For example - Assisting in creating culturally relevant CS lessons that integrate Hawaiian language and place-based knowledge?
  \begin{itemize}
      \item Very positive
      \item Somewhat positive
      \item Neutral
      \item Somewhat Negative
      \item Very negative
  \end{itemize}
  \item How important is it for AI tools to be adapted specifically to support Hawaiian culture and language in educational contexts?
  \begin{itemize}
      \item Extremely important
      \item Very important
      \item Moderately important
      \item Slightly important
      \item Not important at all
  \end{itemize}
  \item Would you be interested in using AI tools that are specifically designed for Hawaiian Immersion education?
  \begin{itemize}
      \item Yes
      \item No
      \item Maybe
  \end{itemize}
  \item How comfortable are you with creating prompts (the instructions you give AI) for AI tools to create teaching materials?
  \begin{itemize}
      \item Extremely comfortable
      \item Very comfortable
      \item Moderately comfortable
      \item Slightly comfortable
      \item Not comfortable at all
  \end{itemize}
  \item What features would be most helpful in an AI tool designed for educators like you? Drop and drop each item to rank from most useful (1) to least useful (8)
  \begin{itemize}
      \item Easy creation of culturally relevant lesson plans
      \item High-quality \textit {`Ōlelo Hawai`i} language support
      \item Integration of place-based knowledge and references
      \item Easy-to-use interface with minimal complexity
      \item Guidance and examples for effective use (e.g., prompt suggestions)
      \item Ability to share and access resources from other teachers
      \item Support for stand-in teachers to maintain continuity
      \item Other (please specify)
  \end{itemize}
  \item In what other ways would you like AI tools or existing tools to be adapted to better meet your needs as an educator? 
  \item Do you have any concerns about AI's role in education that you'd like to share? 
\end{enumerate}
Demographics:
\begin{enumerate}
  \item What is your Ethnic/Racial identity?
  \begin{itemize}
      \item Kanaka Maoli / Native Hawaiian
      \item Other Pacific Islander (please specify)
      \item Asian
      \item Black or African American
      \item White
      \item Hispanic or Latino
      \item American Indian or Alaska Native
      \item Other (please specify)
      \item Prefer not to say
  \end{itemize}
  \item How do you identify in relation to Hawai`i?
  \begin{itemize}
      \item Kanaka Maoli / Native Hawaiian
      \item Kama`āina / Local
      \item Malihini / Recent arrival or visitor
      \item Other (please specify)
  \end{itemize}
  \item What is your age?
  \begin{itemize}
      \item Under 18
      \item 18-24
      \item 25-34
      \item 35-44
      \item 45-54
      \item 55-64
      \item 65 or older
  \end{itemize}
  \item What gender do you identify with?
  \begin{itemize}
      \item Male
      \item Female
      \item Māhū
      \item Non-binary
      \item Prefer not to say
      \item Other (please specify)
  \end{itemize}
\end{enumerate}

\end{document}